\documentclass[twocolumn]{aa}

\usepackage{graphicx}
\usepackage{txfonts}
\usepackage[normalem]{ulem}
\usepackage{xcolor}
\usepackage[utf8]{inputenc}
\definecolor{xlinkcolor}{cmyk}{1,0.6,0,0}

\usepackage[breaklinks=true, 
    bookmarks=false,         
     pdfnewwindow=true,      
     colorlinks=true,    
     linkcolor=xlinkcolor,     
     citecolor=xlinkcolor,     
     filecolor=xlinkcolor,  
     urlcolor=xlinkcolor,      
final=true
]{hyperref}

\begin{document}

   \title{Puzzling Ultra-Diffuse Galaxy Evolution (PUDGE\thanks{The project's playful acronym refers to a character from the video game Dota 2, known for his large size.})}

   \subtitle{I. The existence of a Nube-like galaxy in IllustrisTNG}

   \author{Nata{\v s}a Pavlov\inst{1}
          \and
          Ana Mitra{\v s}inovi{\' c}\inst{2}
          }

   \institute{$^1$ Faculty of Mathematics, University of Belgrade, Studentski trg 16, 11158 Belgrade, Serbia\\
              \email{natasa.pavlov@matf.bg.ac.rs}\\
              $^2$Astronomical Observatory, Volgina 7, 11060 Belgrade, Serbia\\
              \email{amitrasinovic@aob.rs}\\
               }

   \date{Received September 15, 1996; accepted March 16, 1997}

 \abstract{The recent discovery of the most extended ultra-diffuse galaxy (UDG), Nube, has raised yet another question about the validity of the cold dark matter (CDM) model. The studies using cosmological and zoom-in simulations, which assume CDM, failed to replicate galaxies with the structural properties of  Nube. However, the simulation box or the examined population of UDGs may be too narrow to fully capture the range of effects that can lead to the formation of such extraordinary galaxies. In this work we present a case study of a Nube-like galaxy from TNG100, the most extended simulated UDG examined to date that closely mirrors the structural properties of the observed Nube galaxy. Since its formation, the simulated Nube-like galaxy has already been ultra-diffuse and evolved mainly in isolated regions with occasional interactions. Its last major merger was finalized about $1.336$ Gyr ago and left no trace of interaction apart from further extending the stellar size. This evolutionary pathway, featuring a recent merger that expanded an already ultra-diffuse stellar system, is unique and innovative compared to previous studies. We argue that multiple proposed formation mechanisms can operate simultaneously, further expanding the UDGs and making them extreme outliers of the mass-size relation under favorable conditions. Therefore, it is essential to study these simulated extreme outliers, their formation, and, more importantly, their evolution. We also highlight the necessity of carefully analyzing and interpreting the simulated data and better understanding the limitations of a chosen simulation. Thus, if Nube is considered an extreme outlier, its properties are not in tension with the standard cosmological model.} 
  
   \keywords{galaxies: formation --
                galaxies: evolution --
                galaxies: dwarf --
                galaxies: stellar content --
                galaxies: structure -- 
                dark matter
               }

   \maketitle
%

\section{Introduction}
 
Ultra-diffuse galaxies (UDGs) are extremely low-luminosity galaxies with surface brightness $\mu_{V, \textrm{eff}}>25\;\mathrm{mag}\;\mathrm{arcsec}^{-2}$ and effective radius $R_e > 1.5\; \mathrm{kpc}$ \citep{vanDokkum+2015ApJ...798L..45V}. Earlier studies \citep[e.g.,][]{Sandage+Binggeli1984AJ.....89..919S, Impey+1988ApJ...330..634I, Conselice+2003AJ....125...66C}, which focused on the cluster population of UDGs, have found that the UDG subclass comprises galaxies that lack star-forming gas and harbor older populations of stars. Since then, as observational capabilities and techniques have vastly improved \citep{vanDokkum+2015ApJ...798L..45V, Ivezic+2019ApJ...873..111I, Lim+2020ApJ...899...69L, Tanoglidis+2021ApJS..252...18T, Zaritsky+2024AJ....168...69Z}, UDGs have been observed in various environments \citep[for the most complete, up-to-date catalogs of UDGs, see][]{Zaritsky+2023ApJS..267...27Z, Gannon+2024MNRAS.531.1856G}, showing a diversity of fundamental characteristics  (e.g., dark matter and gaseous content, stellar velocity dispersion, number of globular clusters), although not necessarily dependent on the environment.

Studying UDGs, both observationally and theoretically, can shed light on galaxy formation mechanisms and evolutionary paths, probe the nature of dark matter, and test different cosmological models. A recently discovered galaxy named Nube is the most extended massive UDG, seemingly in tension with the cold dark matter (CDM) model \citep{Montes+2024A&A...681A..15M}. The authors showed that the highly extended and flattened structure of Nube is consistent with the fuzzy dark matter model \citep[for a review, see][]{Ferreira2021A&ARv..29....7F}. It was also discussed, based on previous theoretical research \citep[e.g.,][]{DiCintio+2017MNRAS.466L...1D, Chan+2018MNRAS.478..906C, Wright+2021MNRAS.502.5370W, Benavides+2023MNRAS.522.1033B}, that current simulations, which assume the CDM model, do not reproduce UDGs with the structural properties of Nube. Another well-known and puzzling UDG, AGC 114905, was also recently tested against different models \citep{ManceraPina+2024A&A...689A.344M}. Modified Newtonian dynamics was found not to fit the circular velocity of the galaxy. The observed kinematics of AGC 114905 may also be explained by fuzzy dark matter or even self-interacting dark matter \citep[e.g.,][]{Spergel+Steinhardt2000PhRvL..84.3760S, Tulin+Yu2018PhR...730....1T}. The CDM has not been entirely ruled out, although the dark matter halo would require parameters rarely seen in cosmological simulations. 

From a theoretical point of view, a few studies have already explored the fundamental properties and formation pathways of UDGs, utilizing cosmological and zoom-in simulations. \citet{Chan+2018MNRAS.478..906C} showed that the FIRE-2 simulation suite \citep{Hopkins+2018MNRAS.480..800H} can produce UDGs that match observed properties such as surface brightness, effective radius, and stellar mass. In NIHAO simulations \citep{Wang+2015MNRAS.454...83W}, UDGs naturally occur in isolated dwarf-sized halos, due to episodes of gas outflows associated with star formation \citep{DiCintio+2017MNRAS.466L...1D}. Subsequently, \citet{Jiang+2019} explored the formation of UDGs in field and galaxy groups, finding that group UDGs can also form due to tidal puffing up and ram-pressure stripping, suggesting that multiple formation pathways are plausible. Exploring the formation and evolution of UDGs in different environments in the TNG50 simulation \citep[][part of the IllustrisTNG suite]{Nelson+2019MNRAS.490.3234N, Pillepich+2019MNRAS.490.3196P}, \citet{Benavides+2023MNRAS.522.1033B} have concluded that UDGs form through multiple mechanisms, influenced by both intrinsic and extrinsic factors. The variety of origins was also confirmed by \citet{Sales+2020MNRAS.494.1848S}, who focused on the subpopulation of UDGs in galaxy clusters in the TNG100 simulation. Likewise, the UDG cluster subpopulation in the RomulusC zoom-in simulation \citep{Tremmel+2019MNRAS.483.3336T} has different origins, exhibits dynamical support by velocity dispersion, and is in broad agreement with observed galaxies \citep{Tremmel+2020MNRAS.497.2786T}.

The studies mentioned above failed to replicate rare galaxies such as Nube using cosmological and zoom-in simulations, which is often seen as conflicting with the CDM model. However, we must acknowledge the limitations. The simulation box may be too small to capture the full range of environmental effects that can lead to such rare cases, or the examined population of UDGs may be narrow. Given the circumstances, the question is whether Nube can be regarded as an outlier in the UDG population because of its exceptional and extreme nature. The overarching goal of the Puzzling Ultra-Diffuse Galaxy Evolution (PUDGE) project is a detailed examination of the properties and evolution of the simulated UDG population. To do so reliably, we must address the pressing issue that there are observed galaxies seemingly inconsistent with cosmological simulations. This first part of the series focuses on the most puzzling observational example, the extremely extended Nube galaxy. Thus,  we present here a case study of a Nube-like galaxy from TNG100, and we provide insights into how such galaxies may form through a combination of effects, including environmental interactions and specific halo properties. This contributes to the ongoing debate about the nature and origins of UDGs in the Universe, as well as the nature of dark matter in general.

This paper is organized as follows. We describe our methods and candidate selection criteria in Sect.~\ref{sec:methods}. In Sect.~\ref{sec:results} we present the fundamental properties of the galaxy and its dark matter halo, as well as its evolutionary path. In Sect.~\ref{sec:discussion} we discuss the results and implications, considering the previous work. Finally, we provide our concluding remarks and a summary of our work in Sect.~\ref{sec:summary}. 
 
\section{Methods}\label{sec:methods}

The IllustrisTNG\footnote{Publicly available at \url{https://www.tng-project.org/data/}.} cosmological hydrodynamical simulations of galaxy formation \citep{TNGmethods2017,TNGmethods2018,Nelson+2019ComAC} were performed using the \texttt{Arepo} code \citep{Springel2010AREPO} and \citet{PlanckColab+2016} cosmological parameters: matter density $\Omega_\mathrm{m} = 0.3089$, baryon density $\Omega_\mathrm{b} = 0.0486$, dark energy density $\Omega_\Lambda = 0.6911$, and Hubble constant $H_0 = 0.6774\; \mathrm{km}\; \mathrm{s}^{-1}\; \mathrm{Mpc}^{-1}$. In addition to self-consistent modeling of gravitational interactions and magnetohydrodynamical interactions of the gaseous component, the simulations also incorporate processes such as radiative cooling, star and supermassive black hole formation, and their feedback effects, along with chemical enrichment within galaxies \citep[often on a subgrid level][]{TNGmethods2017, TNGmethods2018}.

The simulation suite consists of three runs: TNG50, TNG100, and TNG300, named after the side length of the simulation box ($50$, $100$, and $300$ Mpc, respectively). The boxes differ in cubic volume and particle resolution. In this work we use TNG100 \citep{Marinacci+2018, Naiman+2018, Nelson+2018, Pillepich+2018, Springel+2018}, which perfectly balances volume and resolution. The TNG100 box simulates a larger volume compared to the TNG50 box at a reasonable expense of particle resolution, resulting in a larger sample of galaxies and massive galaxy groups and clusters; low-mass galaxies are not resolved with a large number of particles, which would allow a detailed analysis of their internal structure. The largest volume is simulated in the TNG300 box, which is the best choice for statistical analysis, but the particle resolution imposes a higher mass limit for examining low-mass galaxies. The mass resolution for the TNG100 simulation box is $7.5 \times 10^{6}\;\mathrm{M}_\odot$ for dark matter and $1.4 \times 10^{6}\;\mathrm{M}_\odot$ for baryonic particles. With roughly $250$ particles, a galaxy of stellar mass similar to Nube can have its (stellar) fundamental properties recovered reliably \citep{Onions+2012MNRAS.423.1200O}. However, a more detailed examination of its internal structure typically requires more particles. Specifically, the stellar effective radius, the masses of stellar or gaseous components, and other similar parameters that describe the galaxy as a whole can be treated as reliable, while the internal structure of stars (such as stellar distribution, velocity fields, or stellar orbits) cannot be examined in detail due to the low number of particles.

We imposed the selection criteria based on the observed masses and effective radius to search for Nube-like galaxies in the present-day snapshot, corresponding to $z=0$. The stellar mass is constrained within the range of $8 < \log M_\star \;[\mathrm{M}_\odot] < 9$, and the stellar half-mass radius is within the range of $6 < R_e \;[\mathrm{kpc}] < 8$. However, the gaseous and total dynamical mass is constrained with only the lower limit, $\log M_\mathrm{G} \;[\mathrm{M}_\odot] > 8$ and $\log M_\mathrm{dyn} \;[\mathrm{M}_\odot] > 10$, respectively. This filter yields $23$ candidate galaxies, but only one matches the observed properties of Nube. The remaining candidates have stellar populations embedded in much more massive dark matter halos and gaseous components (see Appendix~\ref{sec:outlier} for a detailed comparison with other galaxies in TNG100). We examine this Nube-like galaxy with a \texttt{SubfindID} 149222 at $z=0$. 

\section{Results}\label{sec:results}

\begin{center}
\begin{table*}[ht!]
\centering
\caption{Fundamental properties of Nube \citep[taken from][]{Montes+2024A&A...681A..15M} and simulated Nube-like galaxy (\texttt{ID149222}) for comparison.} \label{tab:prop}
\begin{tabular}{ccccccccc}
\hline \hline
Galaxy   & log $M_\mathrm{dyn}$ & log $M_\star$ & log $M_\mathrm{HI}$ & $R_e$        & $\Sigma_e$ & $\mu_V(0)$ & Age & $b/a$ \\
 & (log M$_\odot$) & (log M$_\odot$) & (log M$_\odot$) & (kpc) & (M$_\odot$ pc$^{-2}$) & (mag arcsec$^{-2}$) & (Gyr) & \\ \hline
Nube     & 10.42 $\pm$ 0.23   & 8.6 $\pm$ 0.1       & 8.35 $\pm$ 0.12   & 6.9 $\pm$ 0.8 & 0.9 $\pm$ 0.1 & 26.23 $\pm$ 0.07  & 10.2 $^{+2.0}_{-2.5}$ & 0.97 $\pm$ 0.01 \\
\texttt{ID149222} & (10.12, 10.48)          & 8.39            & 8.17          & 6.72      & (0.86, 1.53) & $\sim$ 26.3 &  7.479 & $\sim$ 0.758 \\\hline \hline
\end{tabular}
\end{table*}
\end{center}

We list the fundamental properties of this galaxy in Table~\ref{tab:prop} along with the properties of Nube for comparison. The total gas mass of the Nube-like galaxy is $\mathrm{log}(M_\mathrm{G}\;[\mathrm{M}_\odot])= 8.72$; the total mass of neutral hydrogen, which is lower, is from the supplementary catalog \citep{Diemer+2018ApJS..238...33D, Diemer+2019MNRAS.487.1529D}. Two values are given for the dynamical mass $\log M_\mathrm{dyn}$, where the lower value represents the dynamical mass calculated within $3R_\mathrm{0.5,\star}$ to allow direct comparison with \citet{Montes+2024A&A...681A..15M}, and the higher value represents the total dynamical mass. We also include the average stellar age and axis ratio $b/a$, computed from the raw particle data. Similarly, based on the raw particle data, the stellar metallicity of the simulated Nube-like galaxy is $[\mathrm{Fe/H}]\simeq -3.72$, while the same value for Nube is $[\mathrm{Fe/H}]\simeq -1.09$. These discrepancies between the simulated and observed galaxy are due to the particle resolution. These values obtained from the raw particle data should not be considered reliable enough, given that the Nube-like galaxy is resolved with roughly 250 particles. We also include the central surface brightness in the V filter $\mu_V(0)$ (see Appendix~\ref{sec:mocks} for mock images). Two values are also given for the effective surface density $\Sigma_e$:  the lower limit is calculated with the effective (half-mass) radius of the galaxy, and the upper limit is corrected for projection effects (the effective radius is scaled with $0.75$). There are multiple ways to derive the two-dimensional effective radius from the three-dimensional half-mass radius available in the Subhalo catalog. The simplest solution is to use a scaling factor, as we did here, adopting the optimal value for the scaling factor of \citet{Wellons+2015_075scale}.

The structural parameters of this simulated Nube-like galaxy are, in general, in agreement with the observational values (Table~\ref{tab:prop}), although there are differences in the parameters that depend on the internal structure and how well it is resolved (such as stellar age, axis ratio, or stellar metallicity). Since the internal structure cannot be reliably analyzed with such a low number of stellar particles, we do not expect the galaxy's surface density profile to be as flat as the observed profile. However, we can look at the distribution of its stellar component in two projections (see Appendix~\ref{sec:mocks}) to assess its overall shape and gradient. These two projections do not result in a strikingly different distributions of stars, which would be the case with late-type rotationally supported galaxies (although certain asymmetries are still noticeable). Thus, rotating the galaxy with respect to its stellar angular momentum vector appears to have little effect, indicating that the galaxy is dynamically supported by velocity dispersion (i.e., early-type). Moreover, the galaxy does not exhibit any definitive and distinct tidal features and distortions that could indicate recent interaction, even though the last major merger was completed only $1.336\;\mathrm{Gyr}$ ago, according to the \texttt{SubLink} \citep{Rodriguez-Gomez+2015MNRAS} data.

At present, the galaxy is identified as a part of a massive group of galaxies (mass $M_{200} = 6.68 \cdot 10^{13}\;\mathrm{M}_\odot$ and virial radius $R_{200} = 855.35\;\mathrm{kpc}$), located on the outskirts outside the virial radius, at a distance $D=1696.7\;\mathrm{kpc}$ from the center. It resides in a fairly secluded area, with more massive neighboring galaxies located at distances of $210.67\;\mathrm{kpc}$, $724.25\;\mathrm{kpc}$, and more. Despite the high mass of the host halo, we still recognize it as a galaxy group since, according to \citet{Paul+2017MNRAS.471....2P}, galaxy clusters are groups more massive than $M_{200}=8\cdot 10^{13} \; \mathrm{M_\odot}$.

\begin{figure}[ht!]
\centering \includegraphics[width=\columnwidth, keepaspectratio]{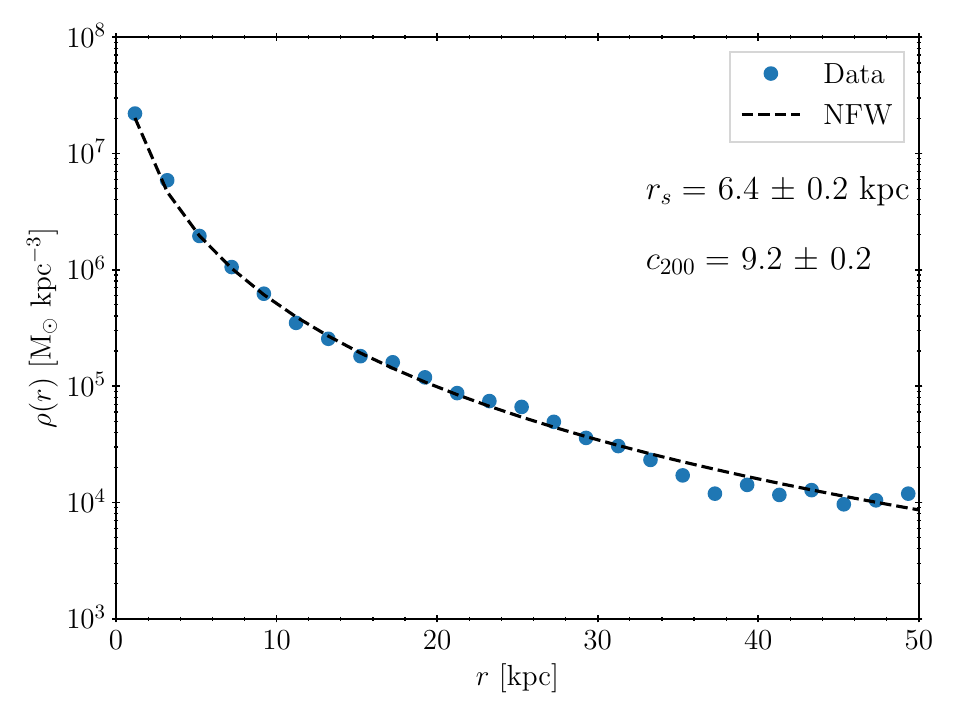}
\caption{Dark matter halo density profile of a simulated Nube-like galaxy, with a NFW profile fit (dashed black line).}
\label{fig:nfwfits}
\end{figure} 

\begin{figure}
\centering \includegraphics[width=\columnwidth, keepaspectratio]{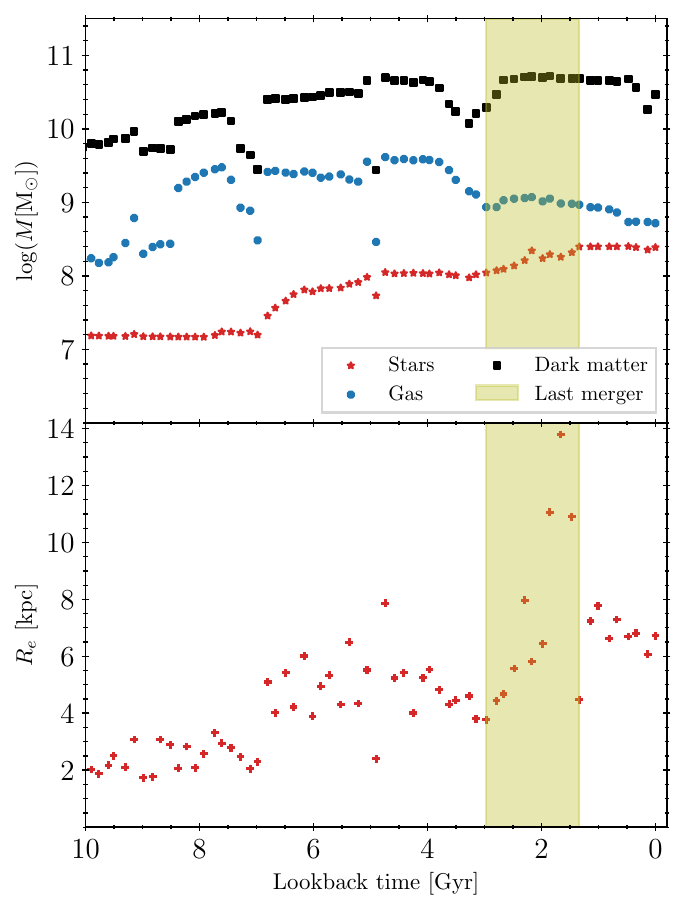}
\caption{Evolution of the galaxy's mass (upper panel) and stellar effective (half-mass) radius (lower panel). The last merger event is shaded.}
\label{fig:evolution}
\end{figure}

The dark matter density profile of the galaxy (Fig.~\ref{fig:nfwfits}) is fitted with the Navarro–Frenk–White (NFW) profile \citep{nfw1997} remarkably well. The scale radius $r_s = 6.4 \pm 0.2\;\mathrm{kpc}$, and the concentration parameter $c_{200} = 9.2 \pm 0.2$ result in a virial radius $R_\mathrm{vir} \simeq 58.88\;\mathrm{kpc}$. We also calculated the spin parameter as defined by \citet{Bullock+spin2001ApJ...555..240B}. The distribution of this parameter is log-normal, with a significant scatter, no mass dependence, and a mean value of $\lambda'\simeq 0.035$. The calculated value for the Nube-like galaxy, $\lambda' = 0.038485$, is very close to the typical one.

To unravel how the stellar component of this galaxy became so extended, we have to look at its history, specifically the evolution of the mass of all the components (dark matter, stars, and gas), as well as the evolution of the stellar half-mass radius (see Fig.~\ref{fig:evolution}). We also marked the time of its last major merger, but instead of marking the moment when the merger was finalized, we marked the whole duration of the merger. The moment when the merger started was chosen based on the mass history of the secondary galaxy: the merger starts when the secondary galaxy is at the peak of its total mass prior to the event, just before the mass transfer between galaxies has started.

During the last $10$ Gyr, the Nube-like galaxy has mainly evolved quietly in isolated regions, with occasional and brief interactions. One of those interactions, roughly $7$ Gyr ago, triggered star formation (practically non-existent prior to that event) that slowly and gradually built up the stellar component of the galaxy. The galaxy was already very extended for its mass, making it ultra-diffuse most likely from its formation (i.e., 10 Gyr ago, the stellar mass was slightly higher than $M_\star \simeq 10^{7} \; \mathrm{M_\odot}$ with $R_\mathrm{e} \geq 2\;\mathrm{kpc}$). As the stellar mass grew gradually, so did its radial extent.

At the beginning of the last merger about $3$ Gyr ago, before mass transfer between the galaxies started, the secondary galaxy had a comparable stellar mass, but a lower total mass than the Nube-like galaxy. The merger can be classified as major even by the mass ratio based on their total masses, which is $\sim 0.36$. The Nube-like galaxy absorbed almost the entirety of the stellar content of the secondary galaxy, and very little star formation has taken place since then. Only about $6\%$ of the total stellar mass (at present) has formed since the merger started. Hence, the enormous gas content of the galaxy did not efficiently turn into stars, and the drop in gas mass we see before merging is most likely reflecting the outer layers being blown away in this violent process. This was to be expected since, before the merger, the gaseous content was very extended, with a half-mass radius well above $20\;\mathrm{kpc}$ and a low average density. Since this is the case, the abrupt increase in the effective radius we see during the merger is due to stellar particles from the secondary galaxy being absorbed, not the stellar content of the host being puffed up initially. The effective radius decreases as these stellar particles slowly settle in the Nube-like galaxy after the merger. However, the resulting host (a present-day Nube-like galaxy) still ends up more diffuse (compared to its state before the merger) and represents the most extended UDG (of the corresponding mass) found in simulations, which closely matches the fundamental properties of Nube (e.g., masses, effective radius).

\section{Discussion}\label{sec:discussion}

The global parameters of this examined galaxy, such as dynamical, stellar, and neutral hydrogen mass, effective radius, effective surface density, and central surface brightness, are in almost perfect agreement with the values derived from observation. However, other structural parameters, which are highly dependent on how well the galaxy is resolved, show discrepancies and should not be considered reliable due to the low number of stellar particles. That does not imply that the inherent limitations of the simulation, such as particle resolution, cannot truly reflect the formation, evolution, and global properties of UDGs. On the contrary, the global properties should be considered reliable \citep{Onions+2012MNRAS.423.1200O}, while only the finer internal structure is prone to inaccuracies. For example, \citet{Sales+2020MNRAS.494.1848S} explored the cluster population of UDGs in the TNG100 box, considering all galaxies with as few as 25 stellar particles.

We mentioned that the examined galaxy is immune to rotation with respect to its stellar angular momentum vector; the shape appears similar in the two projections. This has led us to conclude that the galaxy is an early-type galaxy, supported by velocity dispersion, which is consistent with the fact that most UDGs are dispersion-dominated \citep[e.g.,][]{Sales+2020MNRAS.494.1848S, Tremmel+2020MNRAS.497.2786T}. 

Considering the immediate environment of the Nube-like galaxy, we have found that more massive galaxies are located at distances of $210.67\;\mathrm{kpc}$, $724.25\;\mathrm{kpc}$, and higher. In contrast, the closest massive neighbor to Nube, the galaxy UGC 929, is found to be at a distance of $435\;\mathrm{kpc}$ \citep{Montes+2024A&A...681A..15M}. Although the distances to more massive galaxies are not exactly the same as determined from observations, the secluded immediate environment of our examined galaxy is still in broad agreement. 

\subsection{Post-interaction signatures}

We previously mentioned that the galaxy does not exhibit any distinctive tidal features or distortions despite the recent merger. That does not imply that such features cannot form or that the last merger was so extraordinary that it left no apparent traces. A more likely explanation is that the particle resolution is insufficient to reliably form faint low-mass tidal features. An inherent limitation of any simulation, cosmological or any other, is its particle resolution, which particularly impacts low-mass galaxies. In the low-mass regime, galaxies are resolved with a few hundred stellar particles or fewer (e.g., in the TNG100 box), making a formation of faint low-mass tidal features almost impossible. Furthermore, since those galaxies are not well-resolved, other traces of interactions (e.g., kinematical, disturbed stellar orbits) are of no use because they heavily rely on the number of particles in the system.

Observational efforts also face challenges in the low-mass regime. Detection of tidal features near low-mass galaxies is particularly challenging because of their intrinsically faint luminosities and low surface brightness, which make such features difficult to detect against the background noise. Instrumental improvements can help alleviate these challenges, although they are not the only solution. The application of complex procedures and techniques can also mitigate these issues, as demonstrated by \citet{MMicic+2023ApJ...944..160M}, who applied adaptive smoothing to isolate a faint tidal tail and bridge. Hence, the absence of distinctive traces of recent interactions in the low-mass regime, whether related to simulated or observed galaxies, does not always imply a lack of interactions and could easily be attributed to the limitations mentioned above.

\subsection{Dark matter halo of the galaxy}

We found that the NFW profile \citep{nfw1997} fits the dark matter profile well, with the fitting parameters $r_s = 6.4 \pm 0.2$ and $c_{200} = 9.2 \pm 0.2$. This value of the concentration parameter $c_{200}$ may appear low for a dark matter halo of this mass $\log(M_\mathrm{DM} \;[\mathrm{M}_\odot]) = 10.4714$, as low-mass halos typically have higher concentrations \citep[e.g.,][]{Bullock+profiles2001MNRAS.321..559B, Dutton+Maccio2014MNRAS.441.3359D}. Although it appears lower, this value is still within the expected scatter (at the lower end) and does not represent an extreme outlier that would be impossible to see in simulations.

The fitting parameters result in a viral radius $R_\mathrm{vir} \simeq 58.88\;\mathrm{kpc}$, which differs slightly from the expected virial radius of a dark matter halo with this mass, $\mathrm{log}(M_\mathrm{DM} \;[\mathrm{M}_\odot]) = 10.4714$. Assuming a dark matter halo is virialized, and using the \citet{PlanckColab+2016} data for the value of the critical density of the Universe, we would expect a virial radius $R_\mathrm{vir} \simeq 65.46\;\mathrm{kpc}$. This difference implies that the dark matter halo is not in virial equilibrium, which we additionally confirmed by calculating its kinetic and potential energy. However, this is understandable, considering that the galaxy recently went through a merger.

One of the suggested formation mechanisms is that UDGs can form in higher-spin dark matter halos
\citep[e.g.,][]{Amorisco+Loeb2016MNRAS.459L..51A, Rong+2017MNRAS.470.4231R, Liao+2019MNRAS.490.5182L}. Since we wanted to test whether this holds for the Nube-like galaxy, we calculated the spin parameter as defined by \citet{Bullock+spin2001ApJ...555..240B}, given in Sect.~\ref{sec:results}. The calculated value is close to the typical one, suggesting that the spin is not abnormally high. This is in line with some previous studies that found that the distribution of the halo spin in UDGs is typical \citep{DiCintio+2017MNRAS.466L...1D, Jiang+2019}. Moreover, \citet{Jiang+2019MNRAS.488.4801J} have also found that halo spin is not a good predictor of galaxy size.

\subsection{Implications}

The research of ultra-diffuse galaxies, as extreme examples of low surface brightness galaxies, is in its infancy. Although several formation pathways have been suggested \citep[e.g.,][]{Amorisco+Loeb2016MNRAS.459L..51A, DiCintio+2017MNRAS.466L...1D, Rong+2017MNRAS.470.4231R, Jiang+2019, Liao+2019MNRAS.490.5182L, Tremmel+2019MNRAS.483.3336T, Sales+2020MNRAS.494.1848S, Wright+2021MNRAS.502.5370W, Benavides+2023MNRAS.522.1033B}, there is no complete consensus on the dominant one, due to the diversity of the UDG properties and their environments. To explain this diversity, there is increasing support for the idea that multiple formation pathways are viable, depending on the galaxy's environment and history. The galaxy presented in this work contributes to the ongoing debate with the additional idea that these formation pathways can operate simultaneously: an already ultra-diffuse galaxy can expand its size even further under favorable conditions. This idea should not be considered controversial. For example, after studying observed UDGs, \citet{Fielder+2024AJ....168..212F} could not conclusively dismiss that some galaxies were already formed in the field as UDGs before being processed and puffed up in the group environment.

Although several studies \citep[e.g.,][]{Tremmel+2019MNRAS.483.3336T, Sales+2020MNRAS.494.1848S, Benavides+2023MNRAS.522.1033B} have emphasized the role of external tidal factors in the formation of UDG, \citet{Wright+2021MNRAS.502.5370W} was the only one that reported early mergers as a possible origin. They found that interactions could lower the central density of the gas, disperse it, and compress it in the outer regions of the galaxy, resulting in asymmetric star formation that leads to the formation of UDGs. However, the suggested merger-driven scenario differs from what we observed during the evolution of this galaxy. The merger that occurred here happened quite recently, and the final extent of the galaxy is not a product of asymmetric star formation, but a mass transfer and puffing up. It is also imperative to mention that this event did not leave definitive signs of interaction. This implies that not all mergers leave a distinctive trace that could be used to assess whether a galaxy has had a recent interaction. Tidal features, in general, strongly depend on the characteristics of a preceding interaction and on the geometry of a merger \citep{tt1972};  some interactions are expected to leave no distinguished trace. However, we cannot rule out the possibility that the merger has left undetected signatures simply due to resolution limitations. As discussed previously, there are also challenges in detecting faint tidal features in observations, meaning that interactions (whether mergers or non-mergers) cannot be easily ruled out. Since interactions still occur in isolation, albeit less frequently, even a fairly secluded environment of a galaxy should not be used to dismiss interactions as a possible origin. 

The galaxy we presented here, as a Nube-like case study, not only represents the most extended UDG of its mass found in simulations so far, but also exhibits fundamental properties that closely mirror those of the observed Nube galaxy. This demonstrates that the observed galaxy is not in tension with the currently dominant cosmological model. Furthermore, the evolutionary pathway of this galaxy, with the recent major merger that expanded an already ultra-diffuse stellar system, is fairly unique compared to previous work. First, it broadens our pool of possible UDG formation mechanisms, contributing to the diversity of origins, and can be used to better constrain the diversity of structural and other properties of UDGs. Second, and perhaps more importantly, our finding suggests that UDGs are not the end product of a given galaxy evolution. Instead, we argue that galaxies, even after they reach the ultra-diffuse state, continue to evolve and have the ability to expand their size further, becoming extreme outliers of the mass-size relation \citep[e.g.,][]{Shen+2003MNRAS.343..978S, Du+2024A&A...686A.168D}, given that the favorable conditions for such a scenario are met. These extreme outliers of the mass-size relation can certainly be found in cosmological simulations, although evidently not all of them exhibit the same fundamental and structural properties as Nube. They may also show diversity in their origins and in their evolutionary pathways, further complicating our efforts to fully understand the processes that lead to the formation of UDGs. Some of them may even prove to be the rare result of a series of circumstantial events, and as such they are likely to be found only in larger-volume cosmological simulations \citep[e.g.,][]{Mitrasinovic+2023A&A...680L...1M}. However, exploring the formation and, more importantly, the evolution of the most extreme UDGs in cosmological simulations\footnote{The endeavor that is central to our work in preparation.} can provide invaluable insights and may prove to be the next logical step forward.

Finally, the galaxy presented in this work has significant implications for our understanding of the ecosystem of cosmological simulations. While previous research has suggested that simulations cannot reproduce UDGs as extended as Nube, our finding demonstrates the opposite. This challenges the notion that extreme UDGs are incompatible with the CDM model and highlights the need for careful analysis and interpretation of the simulated data, as well as a better understanding of the limitations of any given simulation.

\section{Conclusions}\label{sec:summary}

The discovery of the most extended ultra-diffuse galaxy so far, named Nube, raised questions about the validity of the CDM model since studies using cosmological and zoom-in simulations, which assume CDM, were not able to reproduce galaxies with such extraordinary characteristics. If Nube represents the rare extreme example of an observed UDG population (and at the moment this does appear to be the case), it would be reasonable to assume that the conditions that led to its formation should be as equally rare. Under that assumption, it becomes understandable why such galaxies were not found in zoom-in simulations or during the explorations of a narrow population of UDGs:  these simulations are not sufficient to capture the full range of effects that can lead to some of the most extreme cases. 

In this work, as a first step in the PUDGE project whose aim is to explore the nature of simulated UDGs, we presented a case study of the most extended galaxy examined in cosmological simulations to date. This galaxy closely mirrors the fundamental properties of   Nube (specifically its masses, effective radius, central surface brightness, and effective surface density), and challenges the notion that such extremely extended galaxies are in tension with the CDM model. We caution against practices when analyzing or interpreting the data generated with simulations that specifically neglect to account for simulated volume and everything it entails (such as a limited sample of galaxies, environments, and events). The simulated Nube-like galaxy has already formed as ultra-diffuse and spent the majority of its lifetime in isolated regions with occasional interactions;  its last major merger interaction was recently finalized, just $1.336$ Gyr ago, which  further expanded the stellar size of the galaxy without leaving other more definitive signs of interaction. Therefore, we argue that the multiple proposed formation mechanisms of UDGs can operate in tandem and produce extreme outliers of the mass-size relation. Such scenarios require favorable evolutionary conditions and are not easily found in zoom-in simulations or when exploring subpopulations of UDGs in larger-volume cosmological simulations (e.g., galaxy cluster subpopulations). Moreover, the merger-driven expansion of the stellar content, particularly this late during the galaxy evolution, was not previously singled out as a possible formation mechanism of UDGs.

The fact that Nube is consistent with the current cosmological model (and can be reproduced in simulations) suggests that we should explore other simulated examples of UDGs that might have been overlooked. This must be done with greater rigor, care, and detail while considering the limitations of any given simulation (whether it is a particle resolution, simulated volume, or other advised cautions). Examining these extreme UDGs will also require a focus on their properties, formation mechanisms, and entire evolutionary pathways, allowing us to gain invaluable insights into the diversity of UDGs.

Ruling out the CDM model will require more observational effort and a significantly higher number of galaxies (detected with state-of-the-art tools and techniques or the upcoming missions) that appear to be in tension with the current models. Until then, we are justified in treating  Nube as an extreme outlier, not a typical ultra-diffuse galaxy.

\begin{acknowledgements}
      We are grateful to the IllustrisTNG team for making their simulations publicly available. The python packages \texttt{matplotlib} \citep{Hunter2007}, \texttt{seaborn} \citep{Waskom2021}, \texttt{numpy} \citep{Harris2020}, \texttt{scipy} \citep{Virtanen2020}, \texttt{pandas} \citep{McKinney2010} and \texttt{pynbody} \citep{pynbody} were used in parts of this analysis. This research was supported by the Ministry of Science, Technological Development, and Innovation of the Republic of Serbia (MSTDIRS) through contract no. 451-03-66/2024-03/200104, made with the Faculty of Mathematics, University of Belgrade, and contract no. 451-03-66/2024-03/200002, made with the Astronomical Observatory (Belgrade, Serbia).
\end{acknowledgements}

%
%

\bibliographystyle{aa}
\bibliography{aa52777-24}

\begin{appendix} 
\section{Mock images}\label{sec:mocks}

\begin{figure}[ht!]
\centering \includegraphics[width=\columnwidth, keepaspectratio]{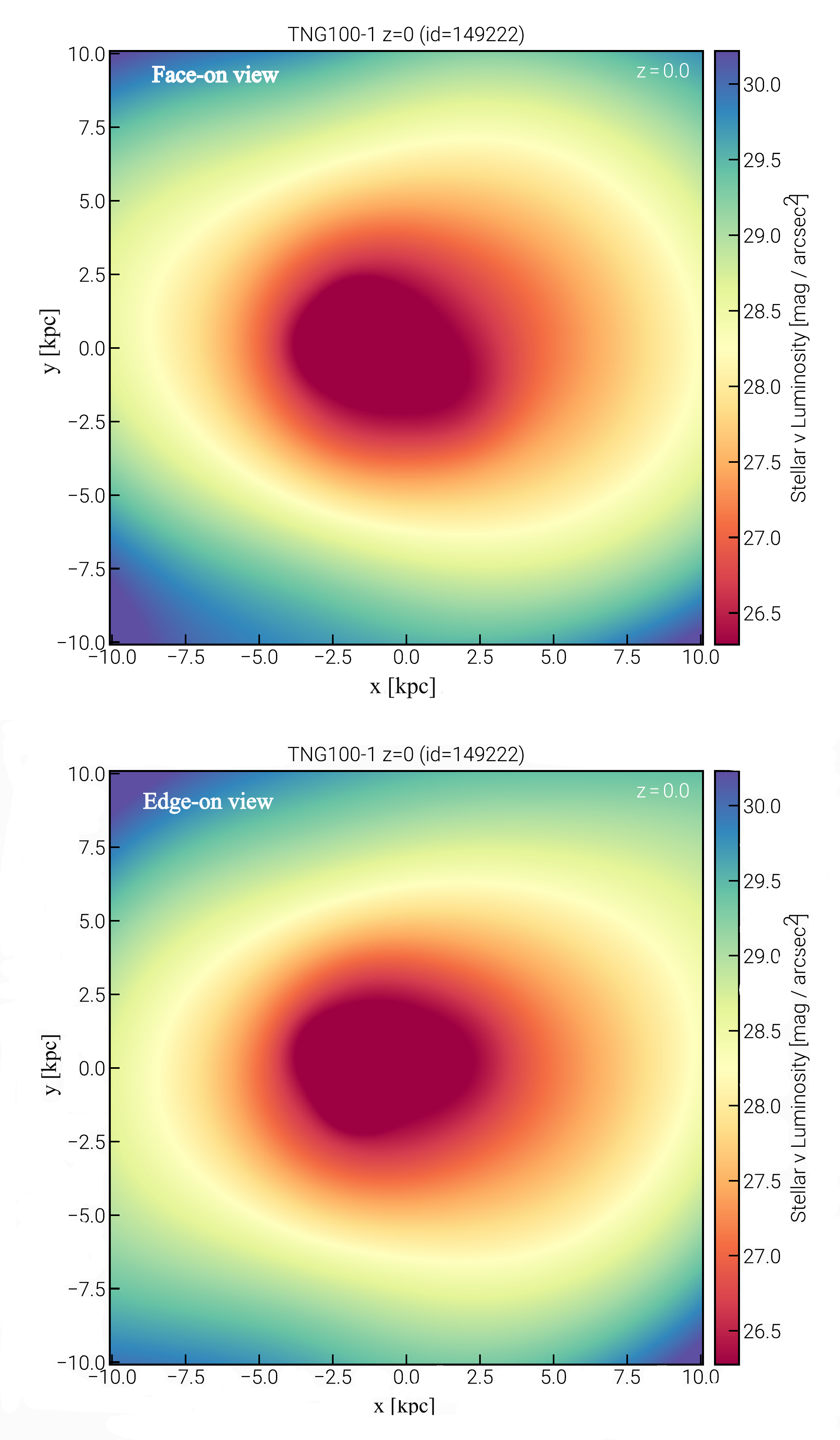}
\caption{Mock images of the Nube-like galaxy in V band, in two different projections: face-on (top panel) and edge-on (bottom panel). Colorbar represents surface brightness in V filter, $\mu_V$, in units that are comparable to observations (i.e., mag$/$arcsec$^{2}$)}
\label{fig:band-v-mocks}
\end{figure}

Figure~\ref{fig:band-v-mocks} was generated using the IllustrisTNG visualization tool\footnote{Available at \url{https://www.tng-project.org/data/vis/}.}. It shows mock images of the examined galaxy in two projections, face-on and edge-on, obtained by rotating the galaxy with respect to its stellar angular momentum vector. The galaxy's shape is discussed in Sect.~\ref{sec:results}.

\section{Comparison of a Nube-like galaxy with other galaxies in TNG100}\label{sec:outlier}

\begin{figure}[ht!]
\centering \includegraphics[width=\columnwidth, keepaspectratio]{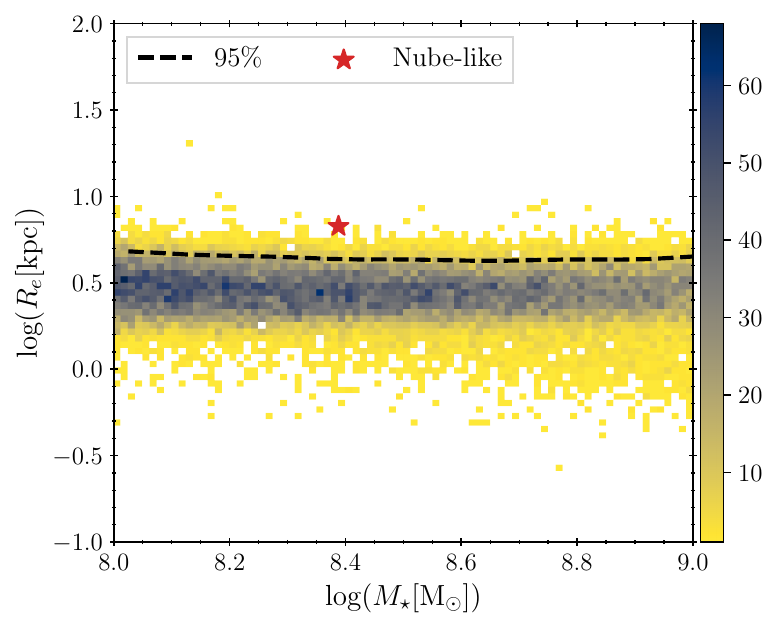}
\caption{Mass-size relation in a narrow stellar mass range, color-coded for the number of galaxies. The Nube-like galaxy is marked, as indicated in the legend. The line (dashed, black) separating the upper 5\% of the most extended galaxies for a given mass is also included.}
\label{fig:mass-size}
\end{figure}

To find a Nube-like galaxy among TNG100 candidates we proceed as follows. First, we excluded spurious self-gravitating baryonic clumps that the subhalo finder occasionally and wrongly identifies as galaxies. That leaves us with a sample of 28486 galaxies in total, which satisfies the initial filtering criterion based on the stellar mass ($8 \leq \log (M_\star) \leq 9$). Examining the galaxies in TNG50, \citet{Benavides+2023MNRAS.522.1033B} have defined UDGs as those that reside in the upper 5\% part of the mass-size relation (i.e., the most extended ones). Here, we adopt the same definition, ending up with a sample of 1440 UDGs. The relevant part of the mass-size relation is shown in Fig.~\ref{fig:mass-size} with the Nube-like galaxy and the critical line that separates the UDG population from the rest of the sample marked. The Nube-like galaxy is evidently very extended for its mass, although there are also other very extended examples.

\begin{figure}[ht!]
\centering \includegraphics[width=\columnwidth, keepaspectratio]{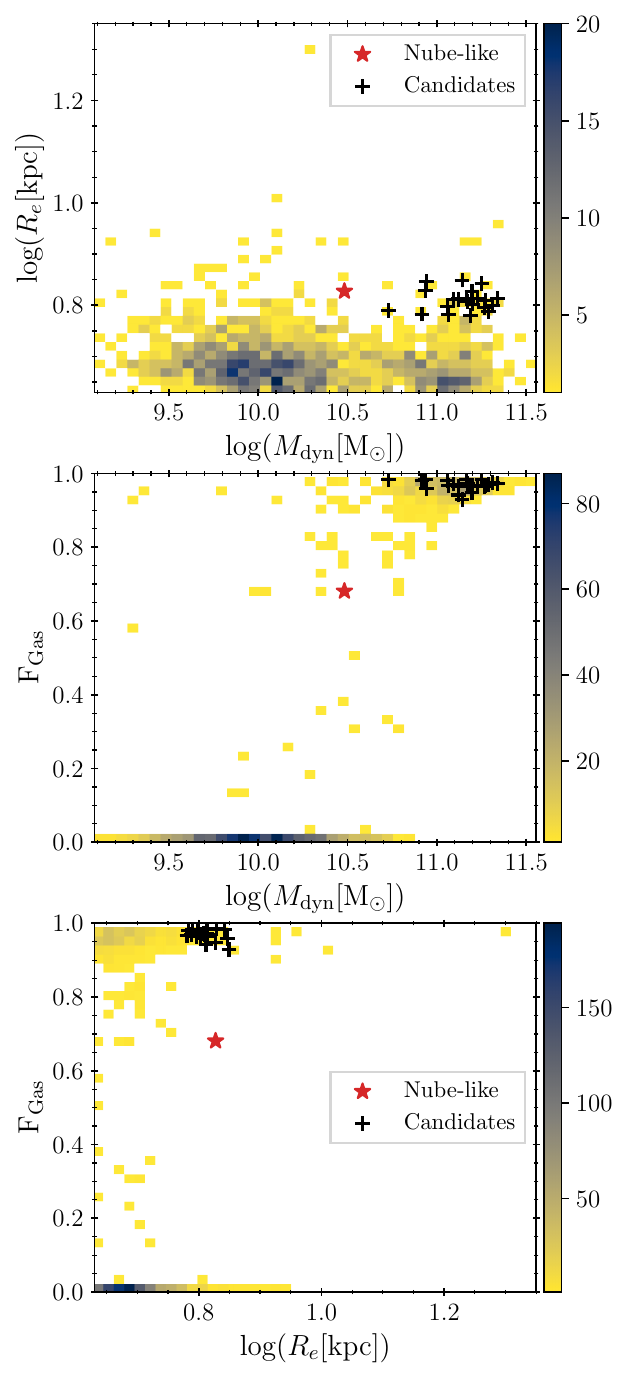}
\caption{Distributions of selected 1440 UDGs in dynamical mass $\log (M_\mathrm{dyn})$, the baryonic gas fraction $\mathrm{F}_\mathrm{Gas}$, and stellar half-mass/effective radius $\log (R_e)$. The legend indicates that the Nube-like galaxy and other candidates are marked separately. The whole UDG sample is color-coded based on the number of galaxies.}
\label{fig:outlier}
\end{figure}

Employing additional filtering criteria (conditions for effective radius, total masses of dark matter and gaseous component) has left us with only 23 candidates, as mentioned in Sect.~\ref{sec:methods}. In Fig.~\ref{fig:outlier} we show two-dimensional distributions of three relevant parameters: total dynamical mass $\log (M_\mathrm{dyn})$, the baryonic gas fraction $\mathrm{F}_\mathrm{Gas}$, and stellar half-mass/effective radius $\log (R_e)$, focusing on a sample of 1440 UDGs and separately marking the Nube-like galaxy examined in this work and other candidates from the curated subsample of 23 galaxies. In the upper panel, $\log (M_\mathrm{dyn})$ -- $\log (R_e)$ distribution, we can clearly notice two distinct \emph{clusters} on the plot: galaxies cluster around $\log (M_\mathrm{dyn})\simeq 10$ and $\log (M_\mathrm{dyn})\simeq 11$, typically on lower effective radii. The most extended galaxies are scattered across all dynamical masses, while the candidates from our subsample are on an extended tail (in terms of effective radius) of the more massive cluster. In contrast, the Nube-like galaxy appears to be an outlier, distinct from both the general UDG population and the rest of the candidates. The unique nature of the Nube-like galaxy is also noticeable when we look at $\log (M_\mathrm{dyn})$ -- $\mathrm{F_{Gas}}$ distribution, where the majority of extremely gas-rich galaxies (as well as all the other candidates) are clustered on higher dynamical masses, and gasless galaxies are typically less massive. Aside from the Nube-like galaxy, other outlier galaxies contain gas but are not extremely gas-rich and have lower masses. Moreover, the Nube-like galaxy is the most extended galaxy that contains gas but is not extremely gas-rich. 

\end{appendix}

\end{document}